\documentclass[twocolumn]{jjap3}
\usepackage{newtxtext,newtxmath}
\usepackage{bm}% bold math
\usepackage{flushend}
\title{The angular dependence of magnetization dynamics induced by a GHz range strain pulse}
\author{Kakeru Tojo, Akira Nagakubo,\thanks{E-mail: nagakubo@prec.eng.osaka-u.ac.jp} and Hirotsugu Ogi}
\inst{Graduate School of Engineering, Osaka University, Suita, Osaka 565-0871, Japan}
\abst{
The dynamics of magnetization is important in spintronics, where the coupling between phonon and magnon attracts much attention. In this work, we study the angular dependence of the coupling between longitudinal-wave phonon and magnon. We investigated  the magnetization dynamics using the time-resolved magneto-optical Kerr effect, which allows measuring spin-wave resonances and the magnetic echo signal. The frequency, mode number, and amplitude of the spin-wave resonance change with the out-of-plane angle of the external magnetic field. The amplitude of the magnetic echo signal caused by the strain pulse also changes with the angle. We calculate these angular dependences based on the Landau-Lifshitz-Gilbert equation and find that the angles of the external field and magnetic moment are important factors for the phonon-magnon coupling when phonon propagates in the thickness direction under the out-of-plane magnetic field.
}

\begin{document}
\maketitle

\section{Introduction}
The interaction between phonon and magnon such as spin-wave\cite{spin_wave}, magnetic resonance\cite{magnetic_resonance}, spin current\cite{spin_current}, spin relaxation\cite{spin_relaxation}, and so on, has attracted great attention. In particular, recent lithography techniques enable us to excite GHz-range surface acoustic waves (SAWs), whose frequencies can be matched to those of the ferromagnetic resonance (FMR). SAW causes the FMR with the same frequency in ferromagnetic film under external magnetic field, which is enhanced at a specific magnetic angle\cite{SAW_angle1, SAW_angle2}. SAW can be focused into a nano region, which enables us to investigate the magnetic properties of single nanomagnets without thermal excitation\cite{SAW_focused}. Recent studies have shown that the smaller the size of nanomagnet, the larger the amplitude of the FMR \cite{SAW_size}. SAW also can ``rewrite'' the magnetization direction recorded in elliptical nanomagnets\cite{SAW_rewritten}.  However, SAW frequency is usually limited to $\sim$10 GHz or less because of the limitation of the interdigital transducer period. 

Longitudinal ultrasound propagating in the out-of-plane direction also can excite FMR. Bayer and coauthors succeeded in exciting 3--40 GHz FMR in a magnetic film deposited on GaAs substrate by a strain pulse launched from the back surface using femtosecond pulse lasers\cite{Longitudinal_FMR1, Longitudinal_FMR2, Longitudinal_FMR3}. A strain pulse excited in the magnetic film also modulates a magnetic response\cite{Strain2magn}, which can be enhanced by acoustic Bragg mirrors\cite{Coupling_BraggMirror}. Sub-THz ultrasound can be used for investigating magnetic properties at ultrahigh frequencies and the interaction between phonon and magnon, which are important to understand the fast magnetic dynamics.

In this study, we generate 10--100 GHz longitudinal ultrasound in a nickel (Ni) thin film using picosecond ultrasonics and observe the magnetic dynamics using the time-resolved magneto-optical Kerr effect (TR-MOKE) measurement. Ultrashort light pulse excites the strain pulse and the spin-wave resonance (SWR) simultaneously. The strain pulse modulates the magnetization dynamics, which largely depends on the angle of the external magnetic field. We measure and calculate the angular dependence of the resonance frequency and amplitudes of magnetization dynamics to evaluate the interaction between ultrasound and magnetization. 

\section{Experiment and Theory}
\subsection{Measurement method}
\begin{figure}
\centering
\includegraphics[width=\linewidth]{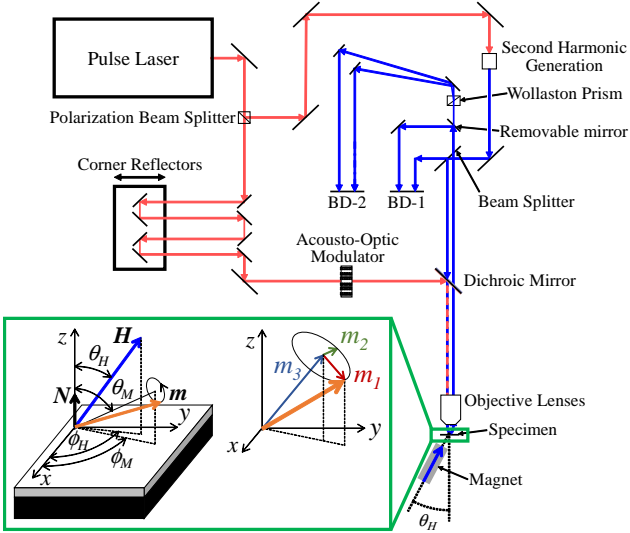}
\caption{Schematic of the optical systems. Red and blue lines denote 800 and 400 nm lights, respectively. BD-1 and BD-2 are used for reflectivity and TR-MOKE measurement, respectively. The inset shows the definition of each coordinate system and parameter.
\label{Chameleon}}
\end{figure}

We excited a broadband strain pulse and SWR by a femtosecond pulse laser. We observe strain and magnetization dynamics by time-resolved reflectivity change\cite{PU_PRL, PU_PRB} and TR-MOKE\cite{TRMOKE1, TRMOKE2} measurements, respectively. A schematic of our developed optics is shown in Fig. \ref{Chameleon}. We applied up to 0.5-T external magnetic field using a permanent magnet by changing its angle $\theta_H$ between 0 and 60 degrees. A strain pulse was excited by an 800 nm pump light, and a linearly-polarized 400 nm probe light was incident perpendicularly on the specimen surface. Details are shown in our previous studies\cite{PU_Elastic-Constant, PU_SrTiO3, PU_isotope_diamond, PU_dielectric_nano-films, USE_PhonMagnCoupling}. We detected the strain pulse from the reflectivity change of the probe light using a balanced detector 1 (BD-1). On the other hand, the magnetization dynamics is detected from the intensity difference between S and P polarization components of the reflected probe light using a balanced detector 2 (BD-2). 

Using the RF magnetron sputtering method, we deposited $\sim$300 nm Ni film capped by 5 nm SiO$_2$. We used a (100) Si substrate whose surface was thermally oxidized. Base pressure, Ar pressure, sputtering power, and substrate temperature during deposition were $\sim$1 $\times 10^{-5}$ Pa, 0.8 Pa, 50 W, and 350 $^\circ$C, respectively. We confirmed that Ni has a polycrystalline structure by X-ray diffraction measurement. 

%Fig{Chameleon}

\subsection{Frequency of spin-wave resonance}
To calculate the frequency of SWR of thin ferromagnetic isotropic film, we consider the following magnetization energy density $E$ normalized by the saturation magnetization $M_s$\cite{SAW_angle2}: 
\begin{eqnarray}
\label{eqenergy}
E & = & -\bm{m} \cdot \bm{H} + \frac{M_s}{2 \mu_0} \left(\bm{m} \cdot \bm{N}\right)^2 - D \Delta \bm{m} \cdot \bm{m},
\end{eqnarray}
where $\bm{m}$ denotes normalized magnetization ($\bm{m} = \bm{M}/M_s$), $\bm{H}$ is the external magnetic field, $\bm{N}$ is a unit vector paralleled to the out-of-plane direction, $D$ is the exchange stiffness, and $\Delta= \frac{\partial^2}{\partial x^2} + \frac{\partial^2}{\partial y^2} + \frac{\partial^2}{\partial z^2}$ is the Laplacian operator. The first, second, and third terms denote the Zeeman, shape anisotropy, and the exchange energy, respectively. 

We calculate the stable angles $\phi^0_{M}$ and $\theta^0_{M}$ through $\frac{\partial E}{\partial \phi_M}=0$ and $\frac{\partial E}{\partial \theta_M}=0$, leading to the following equations: 
\begin{eqnarray}
 \phi^0_{M} & = & \phi_H \\
 \sin { 2 \theta^0_{M}  } & = & \frac{2 \mu_0 H}{M_s} \sin \left(\theta^0_{M} - \theta_H\right)  \label{thetaM}.
\end{eqnarray}
The resonance frequency can be calculated by solving the following Smit-Beljers resonance formula of \cite{FMR-Book, SmitBeljers}
\begin{eqnarray}
 \left(\frac{\omega_{_{SWR}}}{\gamma}\right)^2 = \frac{1}{\sin^{2} \theta^0_{M}} \left\{\frac{\partial^2 E}{\partial \phi_M ^2} \frac{\partial^2 E}{\partial \theta_M ^2}-\left(\frac{\partial^2 E}{\partial \phi_M \partial \theta_M}\right)^2\right\}.\label{Smit_Beljers}
\end{eqnarray}
We obtain the resonance frequency $f_{_{SWR}}=\omega_{_{SWR}} / 2 \pi $ as following\cite{SWR1, SWR2}:
\begin{flalign}
f_{_{SWR}} & = \frac{\gamma}{2 \pi} \sqrt{H_1 H_2}\label{SWRe}\\
H_1 & = H \cos (\theta_H-\theta^0_{M})- \frac{M_s}{\mu_0} \cos^2 \theta^0_{M} + D k_f^2\label{H_1}\\
H_2 & = H \cos (\theta_H-\theta^0_{M})- \frac{M_s}{\mu_0} \cos (2 \theta^0_{M}) + D k_f^2\label{H_2},
\end{flalign}
where $\gamma$ denotes the Gyromagnetic ratio. $k_f$ is the wavenumber vector of spin-wave in the $z$-axis direction, which is defined by a mode number $n$ and the film thickness $L$ as $k_f=\frac{n \pi}{L}$.

\subsection{Amplitude of spin-wave resonance}
We evaluate the amplitudes of SWR. The dynamics of the magnetic moment follows the Landau-Lifshitz-Gilbert (LLG) equation\cite{LLG_LL, LLG_G}:
\begin{eqnarray}
\label{eqLLG}
\frac{\partial \bm{m}}{\partial t} = - \gamma \bm{m} \times \bm{H}_{\text{eff}} + \alpha \bm{m} \times \frac{\partial \bm{m}}{\partial t},
\end{eqnarray}
where $\alpha$ is damping parameter and $\bm{H}_{\text{eff}}$ is the effective magnetic field, which is given by $\bm{H}_{\text{eff}} = -\bm{\nabla} E $ under a static external magnetic field. The SWR is excited by demagnetization $\Delta M_s$ due to instantaneous temperature increase caused by the pump light, which modulates $\bm{H}_{\text{eff}}$ as\cite{SWR_Amp}
\begin{eqnarray}
\bm{H}'_{\text{eff}} & = & \bm{H}_{\text{eff}} + \frac{\partial \bm{H}_{\text{eff}}}{\partial M_s} \Delta M_s \\
& = & \bm{H}_{\text{eff}} - \frac{\Delta M_s}{\mu_0} \cos \theta^0_{M} \bm{\hat{z}},
\end{eqnarray}
where  $\bm{\hat{z}}$ is the $z$-direction unit vector. Assuming that this effective field keeps constant during the resonance period, we can evaluate the change of the magnetic moment to be proportional to the torque of 
\begin{eqnarray}
\gamma \bm{m} \times \bm{H}'_{\text{eff}} &=&\gamma \frac{\Delta M_s}{\mu_0} \cos \theta^0_{M} \sin \theta^0_{M}  \bm{\hat{y}} \\
&=& \gamma \frac{\Delta M_s}{2 \mu_0}\sin { 2 \theta^0_{M}  } \bm{\hat{y}}.
\end{eqnarray}
We can measure the Kerr rotation angle change $\Delta \theta _K$ by the TR-MOKE measurement, which corresponds to the $z$-direction component of the magnetic moment. Therefore, we define the amplitude of SWR as $A_{_{SWR}}$ and evaluate it as follows:
\begin{eqnarray}
A_{_{SWR}} &=&  \left| \gamma \bm{m} \times \bm{H}'_{\text{eff}} \right| \sin \theta^0_{M} \\
 &\propto & \sin { 2 \theta^0_{M}  } \sin \theta^0_{M}.\label{SWR_Amp}
\end{eqnarray}

\subsection{Analysis for the phonon-magnon coupling}
Strain pulse modulates magnetic moment through the magnetoelastic effect. Here, we calculate the dynamics of the magnetic moment and its $z$-component amplitude. By considering the magnetoelastic effect, its energy $E_{_{\text{ME}}}$ is added to $E$:
\begin{flalign}
E &= - \bm{m} \cdot \bm{H} + \frac{M_s}{2 \mu_0} \left( \bm{m}\cdot \bm{N} \right) ^2 -D \Delta \bm{m} \cdot \bm{m} + E_{_{\text{ME}}}\label{Eq_energy}\\
E_{_{\text{ME}}}& = b_1 \left( \varepsilon_{xx} m_{x}^{2} + \varepsilon_{yy} m_{y}^{2} +\varepsilon_{zz} m_{z}^{2} \right) \nonumber \\
& \quad + 2 b_2 \left( \varepsilon_{xy} m_{x} m_{y} + \varepsilon_{yz} m_{y} m_{z} + \varepsilon_{zx} m_{z} m_{x} \right),
\end{flalign}
where $b_1$ and $b_2$ are the magnetoelastic coupling constants. Because we discuss small magnetization modulation from the stable condition, $m_1$ and $m_2$ are much smaller than 1, and magnetic moment in the $m_1m_2m_3$ coordinate system can be written as
\begin{eqnarray}
\bm{m} = \left(
\begin{array}{c}
0  \\
0  \\
1  \\
\end{array}
\right) + \left(
\begin{array}{c}
m_1  \\
m_2  \\
0  \\
\end{array}
\right) + O\left(m_1^2, m_2^2\right)
\end{eqnarray}
as shown in Fig. \ref{Chameleon}. We also assume that displacement $u_z$ and magnetic moment have the forms of $u_z = u_z^0 \exp \left\{ i \left( kz - \omega t \right) \right\}$ and $m_j = m_j^0 \exp \left\{ i \left( kz - \omega t \right) \right\} (j = 1, 2)$, respectively, because we excite the plane wave strain pulse propagating in the $z$ axis. Here, $H_{\text{eff}}$ with respect to the $\bm{m}$ direction is given by
\begin{flalign}
\bm{H}_{\text{eff}}& = -\bm{\nabla}_{\bm{m}} E\\
&= -\left(
\begin{array}{@{}c@{}c@{}c@{}}
\frac{\partial^2E}{\partial m_1^2} m_1 & + & \frac{\partial^2E}{\partial m_1 \partial m_2} m_2  \\
\frac{\partial^2E}{\partial m_1 \partial m_2} m_1 & + & \frac{\partial^2E}{\partial m_2^2} m_2  \\
&\frac{\partial E}{\partial m_3} \\
\end{array}
\right) - \left(
\begin{array}{@{}c@{}}
\frac{\partial E_{_{\text{ME}}}}{\partial m_1}  \\
\frac{\partial E_{_{\text{ME}}}}{\partial m_2}  \\
\frac{\partial E_{_{\text{ME}}}}{\partial m_3}  \\
\end{array}
\right),
\end{flalign}
where $\bm{\nabla}_{\bm{m}} = \left( \frac{\partial}{\partial m_1}, \frac{\partial}{\partial m_2}, \frac{\partial}{\partial m_3} \right)$ is the vector differential operator. We define $h_j= - \frac{\partial E_{_{\text{ME}}}}{\partial m_j}$ and obtain the following relationship from the LLG equation (\ref{eqLLG}):
\begin{flalign}
\left(
\begin{array}{@{}c@{}}
m_1  \\
m_2  \\
\end{array}
\right) & =  \chi \left(
\begin{array}{@{}c@{}}
h_1  \\
h_2  \\
\end{array}
\right)  \label{Eq_m} \\
\chi & =  \frac{1}{C} \left(
\begin{array}{@{}c@{}c@{}}
\frac{\partial^2E}{\partial m_2^2} - \frac{\partial E}{\partial m_3} - \frac{i \omega \alpha}{\gamma} & - \frac{\partial^2E}{\partial m_1 \partial m_2} - \frac{i \omega}{\gamma} \\
- \frac{\partial^2E}{\partial m_1 \partial m_2} + \frac{i \omega}{\gamma} & \frac{\partial^2E}{\partial m_1^2} - \frac{\partial E}{\partial m_3} - \frac{i \omega \alpha}{\gamma} \\
\end{array}
\right)\\
C &= \left(\frac{\partial^2E}{\partial m_1^2} - \frac{\partial E}{\partial m_3} - \frac{i \omega \alpha}{\gamma} \right) \left(\frac{\partial^2E}{\partial m_2^2} - \frac{\partial E}{\partial m_3} - \frac{i \omega \alpha}{\gamma} \right)\nonumber \\
 &\quad - \left(\frac{\partial^2E}{\partial m_1 \partial m_2}\right)^2 - \left(\frac{\omega}{\gamma}\right)^2.
\end{flalign}
The measurable amplitude $A_{pm}$ induced by the phonon-magnon coupling is the $z$-direction component of the magnetization dynamics. We assume that the magnetization dynamics is caused by longitudinal phonons of $u_z = u_z^0 \exp \left\{ i \left( kz - \omega t \right) \right\}$, resulting in only non-zero strain of $\varepsilon_{zz}$. Therefore, by substituting $E$ in Eq. (15) into $m_1$ in Eq. (20), $A_{pm}$ can be written as follows:
\setcounter{equation}{22}
\begin{flalign}
A_{pm} &= m_1^0 \sin \theta^0_{M} \\
&\propto \frac{ \sqrt{H_2^2 + \left( \frac{\alpha \omega}{\gamma}\right)^2 }} { \sqrt{a_0 + a_2 \left(\frac{\omega}{\gamma}\right)^2  + a_4 \left(\frac{\omega}{\gamma}\right)^4 }} \sin { 2 \theta^0_{M}  } \sin \theta^0_{M}\label{Eq_Coupling} \\
a_0 & = H_1^2 H_2^2\\
a_2 & = \alpha^2 \left(H_1^2+H_2^2\right) - 2 H_1 H_2 \\
a_4 & = \alpha^2+1,
\end{flalign}
where $H_1$ and $H_2$ are functions of $\theta_H$ and are calculated by Eqs. (6) and (7), respectively. 
In Eq. (24), we omitted $b_1$ and $\varepsilon_{zz}$ because they are independent from $\theta^0_{M}$ and $\omega$.

\section{Results and Discussion}

%Fig{PU}
\begin{figure}[b]
\includegraphics[width=\linewidth]{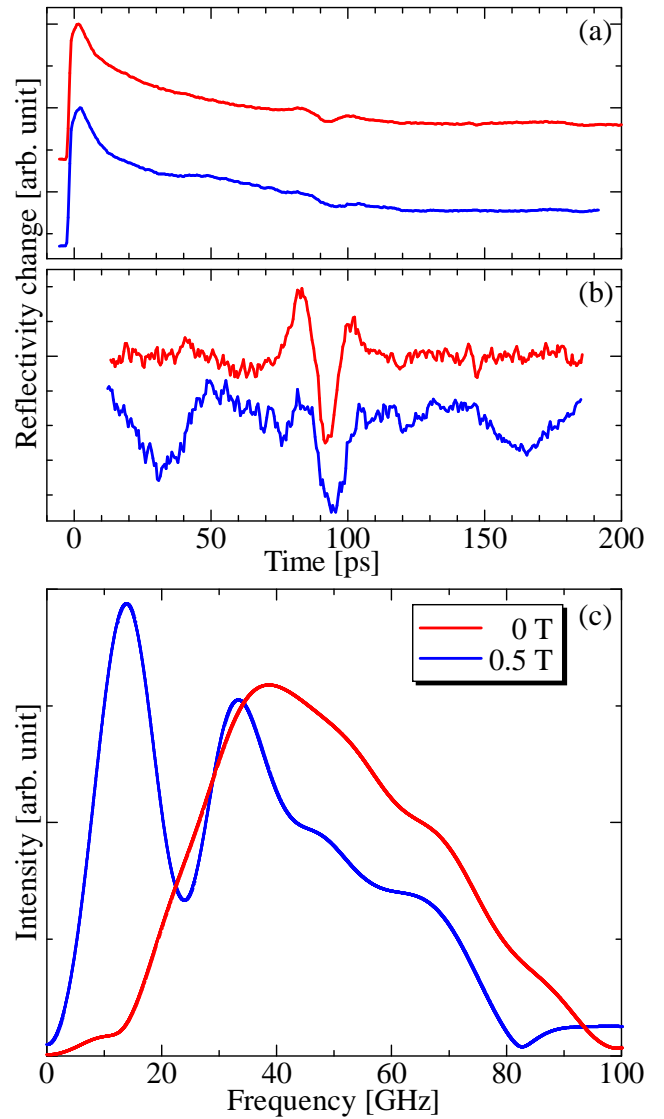}
\centering
\caption{
(a) Observed reflectivity changes, (b) extracted signals, and (c) corresponding FFT spectra. Red and blue lines show the results under no and 0.5 T magnetic fields, respectively. 
\label{PU}}
\end{figure}

At first, we observed propagation of the strain pulse from the reflectivity measurement. The reflectivity change represents the electron-density change caused by strain. Figure \ref{PU}(a) shows measured reflectivity changes, which reflect fast electron diffusion, slower thermal diffusion, and strain-pulse echoes. We extract the echo signals as shown in Fig. \ref{PU}(b), where the background changes of thermal diffusion are subtracted by fitting polynomial functions. Then, we evaluate each frequency amplitude by applying the fast Fourier transformation (FFT) as shown in Fig. \ref{PU}(c). We observed echo signals every 95.4 ps, which leads to the thickness $L$ of $\sim$280 nm from the velocity of 5.8 nm/ps for polycrystalline Ni\cite{Ni_elastic_constant}. We also observed reflectivity changes under the 0.5-T magnetic field at the angle $\theta_H$ of 20 degrees. Under the magnetic field, background changes contain not only thermal diffusion but also $\sim$10 GHz small oscillations as shown in the corresponding FFT spectra of Fig. \ref{PU}(c). The appearance of strain oscillations suggests the magnetostriction effect due to SWR. 

%Fig{Kerr}
\begin{figure}[t]
\includegraphics[width=\linewidth]{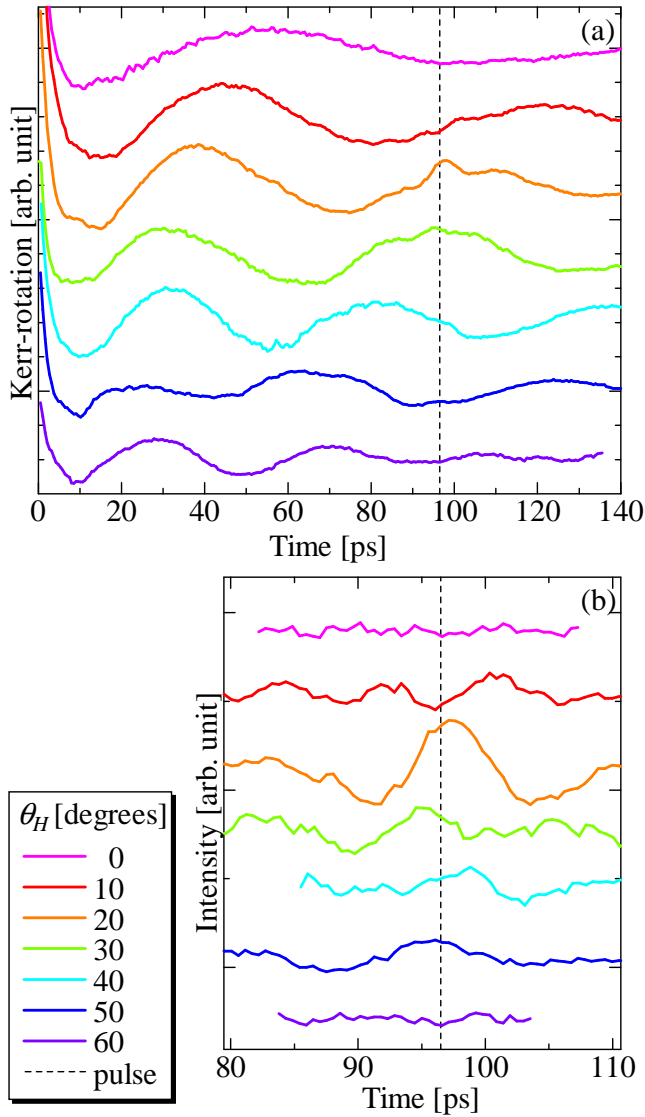}
\centering
\caption{(a) Observed Kerr-rotation angle changes and (b) extracted echo signals by changing the angle of the external magnetic field $\theta_H$. 
\label{Kerr}}
\end{figure}

\begin{figure}
\includegraphics[width=\linewidth]{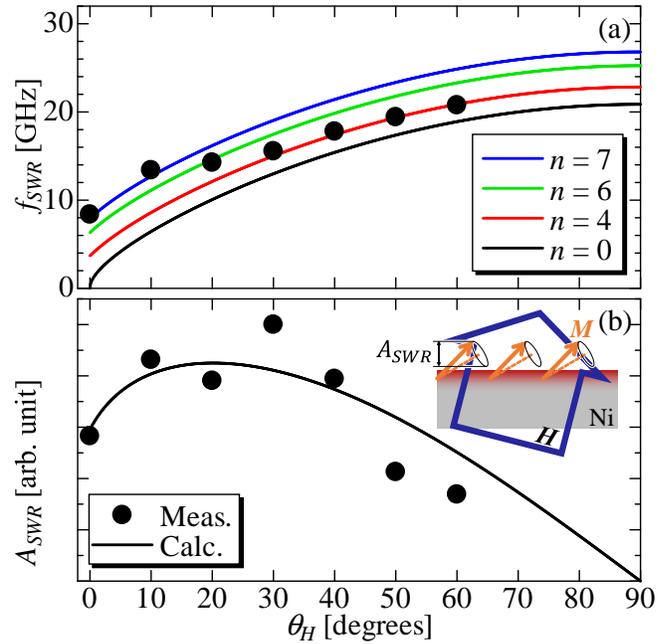}
\centering
\caption{The dependence of (a) SWR frequencies $f_{_{SWR}}$ and (b) SWR amplitudes $A_{_{SWR}}$ on $\theta_H$. Black circles and solid lines denote measured and calculated values. 
\label{SWR}}
\end{figure}

Using TR-MOKE measurement, we observed the magnetization dynamics by changing the angle of the external field $\theta_H$ between 0 and 60 degrees. The observed SWRs are shown in Fig. \ref{Kerr}(a), where background changes are subtracted by fitting polynomial functions. We apply FFT to determine their frequencies and also evaluate the amplitudes from the peak-to-peak heights of the SWR responses. The dependences of them on $\theta_H$ are shown in Figs. \ref{SWR}(a) and (b). We observed 8--21 GHz SWR, and its frequency increases with the increase in $\theta_H$. The measured SWR frequencies agree with the calculations from Eq. (\ref{SWRe}) for the 7th-mode SWR at low $\theta_H$, and they approach those of the 4th-mode SWR at higher angles as shown in Fig. \ref{SWR}(a). In the calculation, we used $D = 0.157$ nm A\cite{kittel}, $M_s = 0.61$ T\cite{kittel}, $\gamma=35.2$ kHz m/A with $\mu_0 H = 0.5$ T and $L = 276$ nm. The higher modes are excited by the local and large initial changes of magnetic moment angle $\theta_M$: The penetration depth of the 800-nm pump light for Ni is $\sim$13.0 nm\cite{Ni_extinction_coefficient, Calculating_extinction_coefficient}, which causes the demagnetization near the surface due to temperature increase and excites the higher-mode SWR. The lower $\theta_H$ also contributes to the larger initial change of $\theta_M$ because $\theta_M$ is highly affected by the demagnetizing field due to shape anisotropy at lower $\theta_H$. We calculate the  changes of $\theta_M$ due to the decrease in $M_s$ from Eq. (\ref{thetaM}), which increases with the decrease in $\theta_H$. The dependence of measured amplitudes of SWR on $\theta _H$ agrees with that of calculation using Eq. (\ref{SWR_Amp}) as shown in Fig. \ref{SWR}(b). These results ensure the validity of our measurement and calculation. Torque caused by the initial thermal strain should have the dependence on $\theta_H$ as discussed in the following section, however, it will be much smaller than the thermal demagnetization and will not affect the amplitude of SWR. 

%Fig{SWR}

\begin{figure}
\includegraphics[width=\linewidth]{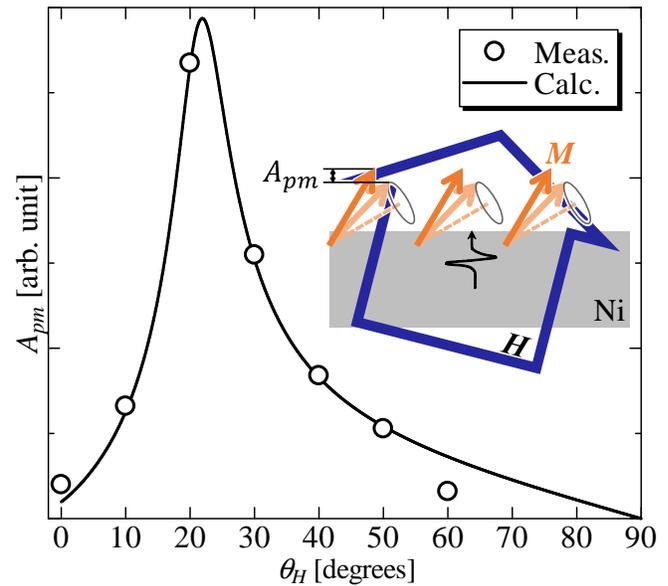}
\centering
\caption{The dependence of measured pulse signal amplitudes on $\theta_H$. Black circles and solid line denote measured $A_{pm}$ and calculated $A_{pm}$, respectively.
\label{Coupling}}
\end{figure}

We also observed a pulse signal around 95 ps in the TR-MOKE measurement, and extract it by subtracting other responses using a polynomial function as shown in Fig. \ref{Kerr}(b). We determine its amplitude $A_{pm}$ from the peak-to-peak height, whose dependence on $\theta_H$ is shown in Fig. \ref{Coupling}. The amplitude becomes the largest at $\theta_H=20$ degrees. This pulse insists that the coherent strain pulse causes the changes in a magnetic moment and the interaction has a large dependence on $\theta_H$. Calculated $A_{pm}$ from Eq. (\ref{Eq_Coupling}) agrees well with the measured amplitudes for $f=10.7$ GHz and $\alpha=0.1$, indicating that the coupling effect can be increased by adjusting the angle as well as the intensity of the external magnetic field. This dependence stems from the change of the torque caused by the strain: According to the LLG equation (\ref{eqLLG}), the torque becomes smaller at lower $\theta_M$ because $\bm{m}$ and $\bm{h}$ are parallel. On the other hand, at higher $\theta_M$, $\bm{h}$ becomes smaller because the initial $z$ component of $\bm{m}$ becomes smaller, resulting in vanishment of the torque. Therefore, the torque caused by the $z$-direction magnetoelastic field takes a maximum at a specific angle. Similar angle dependence is reported by the in-plane SAW-FMR coupling\cite{SAW_angle1, SAW_angle2}. In the out-of-plane magnetic fields, the angle and its intensity affect magnetic moment much more because of the demagnetization field. These results suggest that the SWR at 8--21 GHz is excited by thermal demagnetization at 0 ps, and the magnetization oscillation is excited by phonon-magnon coupling at $\sim$95 ps when the propagated strain pulse returns near the surface. Our measurement and calculation reveal that the $\theta_H$ dependence highly affects the phonon-magnon interaction in the out-of-plane field, which contributes to the investigation of the phonon-magnon coupling in a higher frequency region.

%Fig{Coupling}

\section{Conclusions}
We excited the strain pulse and higher-mode spin-wave resonances by a femtosecond pulse laser. Measured frequencies and amplitudes of the spin-wave resonances agree with the theoretical calculation, ensuring the validity of our measurement and calculation. We also observed the magnetic response caused by the strain pulse, which has a large dependence on the angle of the applied external magnetic field. We experimentally and analytically clarify that the coupling effect can be maximized at certain angles. In addition, we find the contribution of a certain frequency component of the strain pulse to the coupling. Our results will contribute to the measurement of the coupling between longitudinal phonons and magnons that becomes important for higher-frequency techniques.

\acknowledgment
This work was supported by KAKENHI Grant No. 18H01859 of Grant-in-Aid for Scientific Research (B).

\end{document}